\documentclass[12pt]{article}
\usepackage{epsfig}
\usepackage{amsmath}
\usepackage{amssymb}
\usepackage{graphicx}
\usepackage{lscape}
\usepackage {subfigure}
\usepackage{hyperref, cite}
\def\be{\begin{equation}}
\def\ee{\end{equation}}
\def\ba{\begin{array}}
\def\ea{\end{array}}
\def\beqn{\begin{eqnarray}}
\def\eeqn{\end{eqnarray}}

\def\bt{\begin{tabular}}
\def\et{\end{tabular}}
\def\bc{\begin{center}}
\def\ec{\end{center}}

\begin{document}
\title{Testing Texture Two Zero Neutrino Mass Matrices Under Current Experimental Scenario}
\author{Madan Singh$^{*} $\\
\it Department of Physics, M.N.S Government College Bhiwani,\\
\it Haryana,127021, India.\\
\it $^{*}$singhmadan179@gmail.com
}
\maketitle
\begin{abstract}

The latest data from Planck collaboration has presented an improved sensitivity limit of the sum of neutrino masses ($\Sigma$), $\Sigma <0.17eV$ at 95$\%$ confidence level (CL). On the other hand, the updated global fits of neutrino oscillation have shown a refined range of atmospheric mixing angle $\theta_{23}$ at the same CL.  In the light of these observations, we have re-investigated the five viable cases ( $B_{1, 2, 3, 4}$ and  $C$) of texture two zero Majorana mass matrix in the flavor basis. Using the present Planck's data, we have demonstrated that only cases $B_{2}$ and $B_{4}$ are now viable for normal mass ordering, while remaining three cases does not meet the present experimental constraints at 2$\sigma$ CL. The phenomenological predictions of viable cases are also found to be in tune with the latest T2K, Super-Kamiokande and NO$\nu$A results, showing the preference for  normal neutrino mass ordering ($m_{1} < m_{2}<m_{3}$), a maximal Dirac CP-violating phase ($\delta \simeq 270^{0}$), and upper octant of neutrino mixing angle ($\theta_{23} > 45^{0}$). In addition, the implication of $\Sigma$ on neutrinoless double beta decay is studied for viable cases.
\end{abstract}

\section{Introduction}
 
Neutrino physics, at present, is going through the precision era as far as the physical parameters are concerned. In this regard, some notable progress has been seen in the recent past years, which triggers optimism in the lepton sector at significantly higher confidence level(CL). Thanks to three leading reactor neutrino experiments, Daya Bay, Double Chooz and RENO, the smallest mixing angle $\theta_{13}$, also known as, reactor angle, is measured at a precision level of 10$\%$ \cite{1,2,3,4,5}, which, in turn, not only open up the possibilities to explore the leptonic CP violation, but also helpful in pinning down the  octant of atmospheric mixing angle $\theta_{23}$ \cite{6}(which may be either $\theta_{23}> 45^{0}$ or $\theta_{23}< 45^{0}$ or even maximal i.e. $\theta_{23}= 45^{0}$).  The results obtained from T2K, Superkomiokande and  NO$\nu$A experiments  as well as the global fit analyses [see ref.\cite{7, 8, 9, 10,11, 13, 14}], in the recent past years,  have   shown  some significant improvements regarding the Dirac CP violating phase ($\delta$), which is the measure of leptonic Dirac CP violation. The up-to-date data available from these experiments favors the maximal CP violation, $|sin \delta|$ $\sim$ 1, with a significant preference for $sin\delta < 0$. Referring to the updated global analysis \cite{10}, $\delta$ is found to be constrained within $\sim 15 \%$ ($\sim 9 \%$) uncertainty in the normal mass ordering (NO) (inverted mass ordering (IO)) around near maximal CP violating values, $\delta\simeq 270^ {0} $, while $\delta\simeq 90^ {0} $ is now excluded at more than 4$\sigma$ CL. On the other hand, the latest result obtained  from  NO$\nu$A experiment also provides the strong indications regarding the octant of $\theta_{23}$,  excluding the maximal value of $\theta_{23}$, i.e.  $\theta_{23} = 45^{0}$, at  2.6$\sigma$ CL \cite{7}. The  global fit analyses of neutrino oscillation reported in \cite{8, 9,10}, indicate a slight preference for the upper octant ($\theta_{23}> 45^{0})$  at $<2\sigma$ CL, although both the octants are still favored at 3$\sigma$ CL.  In addition, for the first time, a considerable preference for NO over the IO has been reported  at 3$\sigma$ CL \cite{9,10}, with the help of a coherent contribution from various data sets. 
The statistical data basically come from long-baseline accelerator experiments, and their interplay
with short-baseline reactor experiments, where mass-ordering effects can be understood in terms of $\theta_{13}$ . The new data from T2K and NO$\nu$A experiments, possibly combined by the collaborations
themselves \cite{12}, will further test the current trend favoring NO over IO in the present data sample.\\
Apart from the significant progress shown regarding the neutrino oscillation, a new  cosmological data from Planck collaboration \cite{15} has presented a refined limit, 
 \begin{equation} \label{eq1}
  \Sigma\equiv m_{1}+m_{2}+m_{3}< 0.17 \textrm{eV ~~   (at ~~ 95 \% CL}),
 \end{equation}
 on sum of the neutrino masses ($\Sigma$). The above limit has been obtained in the framework of three degenerate neutrino masses, and a $\Lambda$CDM model. The analysis  has been  done using the combination of the Planck temperature power spectrum with Planck polarization, and the Baryon Acoustic Oscillation data.  On the other hand,  KamLAND-Zen experiment has found an improved limit for neutrinoless double-beta ($0\nu\beta\beta$) decay  \cite{16}, as  $|M_{ee}|$ $ <(0.061-0.165) eV$ at 90$\%$ (or $<2\sigma$) CL, where
 \begin{equation}\label{eq2}
|M_{ee}|=|m_{1}c_{12}^{2}c_{13}^{2}e^{2i\rho}+m_{2}s_{12}^{2}c_{13}^{2}e^{2i\sigma}+m_{3}s_{13}^{2}|.
\end{equation}
 The effective mass, in Eq. \ref{eq2}, is just the absolute value of $M_{ee}$ component of the neutrino mass matrix. The observation of $0\nu\beta\beta$ decay, in future, would determine the Majorana nature of neutrinos. For recent searches on $0\nu\beta\beta$ decay see Refs.\cite{16,17,18,19, 20}.
 The sensitivities of these searches correspond
to mass scales in the so-called quasi-degenerate mass region. Further improvement
will allow $|M_{ee}|$ to be probed below 50 meV, starting to constrain the inverted mass hierarchy region under the assumption that neutrinos are Majorana particles.

The above statistical datas from different neutrino related experiments have been taken at considerably higher CL, and hence appears to be promising as far as exploring the flavor structure of neutrino mass matrix is concerned. The construction of mass matrix is necessary for model building, which, in turn, may unravel the underlying dynamics of neutrino masses, mixing and CP violation.  In this context, the texture zero approach has been widely followed in the literature. In particular, texture two zeros have been relatively more successful in both flavor as well as in non flavor basis \cite{21,22,23,24,25,26, 27}.  Apart from  higher predicting power  among the textures zeros, two zero mass matrices, in the flavor basis, can easily result from
underlying flavor symmetries \cite{28,29} and, in addition,  can be realized within the framework of
the see-saw mechanism \cite{29,30}. Moreover, they can, also, be obtained in the context of GUTs
based on SO(10) \cite{31}. Thus, the study of texture two zeros  is highly motivated on theoretical fronts.
 
The present neutrino oscillation data allows only seven out of total fifteen cases of texture two zero mass matrices at 3$\sigma$ CL, as also shown in earlier Refs. \cite{23,24,25,26,27}. Out of these seven cases,  $A_{1}$ and $A_{2}$ predict $|M_{ee}|=0$,  while $B_{1,2,3,4}$ and C predict non-zero $|M_{ee}|$, for the neutrinoless double-beta decay.  Therefore, the present analysis is restricted to following cases,
 \begin{small}
 \begin{equation*} 
B_{1}:\left(
\begin{array}{ccc}
    \times& \times & 0 \\
  \times & 0 & \times\\
  0& \times & \times \\
\end{array}
\right), B_{2}:\left(
\begin{array}{ccc}
    \times& 0 &\times \\
  0 & \times & \times\\
  \times& \times & 0 \\
\end{array}
\right), \\
\end{equation*}
\begin{equation}\label{eq3}
B_{3}:\left(
\begin{array}{ccc}
    \times& 0 &\times \\
  0 & 0 & \times\\
  \times& \times & \times \\
  \end{array}
\right),
  B_{4}:\left(
\begin{array}{ccc}
    \times& \times &0 \\
  \times & \times & \times\\
  0& \times & 0 \\
\end{array}
\right);
\end{equation}
\begin{equation} \label{eq4}
C:\left(
\begin{array}{ccc}
    \times& \times & \times\\
  \times & 0 & \times\\
  \times& \times & 0 \\
\end{array}
\right),
\end{equation}
\end{small} 

in order to study the implication  for neutrinoless double beta decay.  The  nomenclature has been used from  Ref. \cite{23}. Using the experimental constraints, it is found in the earlier analysis  \cite{23,24} that all the five cases predict the quasi degenerate neutrino mass spectrum. 

The purpose of the present work is to upgrade the analysis of D. Meloni et al. \cite{25}  in the light of latest neutrino oscillation data, Planck collaboration data as well as KamLAND-Zen Data.   Earlier, D. Meloni et al. have carried out a detailed analysis of five  cases ($B_{1, 2, 3, 4}$, C) pertaining to  two zero Majorana mass matrix at 1$\sigma$ CL. However, the improved limit of Planck's satellite data as well as the updated global fit analysis of neutrino oscillation warrant the re-examination of texture two zero cases. Among the five cases, we find that only two ($B_ {2} $, and $B_ {4} $), for NO, are now compatible with the new data at 2$\sigma$ CL, while remaining cases are now ruled out at the same level. To this end, we have explicitly shown the incompatibility of the cases through the correlation between $\Sigma$ and $\theta_{23}$. In comparison with the analysis of D. Meloni \cite{25}, our predictions remain valid at relatively higher CL. In addition, the phenomenological predictions for the viable cases are found to be in good agreement with the  results obtained from  T2K, Super-Kamiokande, NO$\nu$A as well as KamLAND-Zen experiments.  Earlier in Ref. \cite{22}, R. Verma  has carried out the analysis for Fritzsch-like texture four-zero lepton mass matrices, in the light of these experimental observations.

 The rest of the analysis is organized as follows: In Section 2, we discuss the methodology used
to reconstruct the neutrino mass matrix and texture two zero conditions. In Section 3, we present the numerical analysis using some analytical relations and correlation plots. In Section 4, we summarize and conclude our work.
 
 \begin{table}[hb]
\begin{small}
\begin{center}
\resizebox{12cm}{!}{
\begin{tabular}{|c|c|c|c|c|}
\hline
Parameter& Best Fit & 1$\sigma$ & 2$\sigma$ &
3$\sigma$ \\ \hline $\delta m^{2}$
$[10^{-5}eV^{2}]$ & $7.55$& $7.39$ - $7.75$ &
$7.20$ - $7.94$ & $7.05$ - $8.14$ \\ \hline
$|\Delta m^{2}_{31}|$ $[10^{-3}eV^{2}]$ (NO) &
$2.50$ & $2.47$ - $2.53$ & $2.44$ - $2.57$ &
$2.41$ - $2.60$\\ \hline $|\Delta m^{2}_{31}|$
$[10^{-3}eV^{2}]$ (IO) & $2.42$ & $2.38$ - $2.45$
& $2.34$ - $2.47$ & $2.31$ - $2.51$ \\ \hline
$\theta_{12}$ & $34.5^{\circ}$ & $33.5^{\circ}$ -
$35.7^{\circ}$ & $32.5^{\circ}$ - $36.8^{\circ}$
& $31.5^{\circ}$ - $38^{\circ}$\\ \hline $
\theta_{23}$ (NO) & $47.7^{\circ}$
&$46^{\circ}$ - $48.9^{\circ}$  &
$43.1^{\circ}$ - $49.8^{\circ}$ & $41.8^{\circ}$
- $50.7^{\circ}$ \\ \hline $\theta_{23}$ (IO)&
$47.9^{\circ}$ & $46.2^{\circ}$ - $48.9^{\circ}$
& $44.5^{\circ}$ - $48.9^{\circ} $&
$42.3^{\circ}$ - $50.7^{\circ}$ \\ \hline
$\theta_{13}$ (NO) & $8.45^{\circ}$ &
$8.31^{\circ}$ - $8.61^{\circ}$ & $8.2^{\circ}$ -
$8.8^{\circ}$& $8.0^{\circ}$ - $8.9^{\circ}$ \\
\hline $\theta_{13}$ (IO) & $8.53^{\circ}$ &
$8.38^{\circ}$ - $8.67^{\circ}$ & $8.3^{\circ}$ -
$8.8^{\circ}$ & $8.1^{\circ}$ - $9.0^{\circ}$ \\
\hline $\delta$ (NO) & $218^{\circ}$ &
$191^{\circ}$ - $256^{\circ}$& $182^{\circ}$ -
$315^{\circ}$ & $157^{\circ}$ - $349^{\circ}$ \\
\hline $\delta$ (IO) &$281^{\circ}$&
$254^{\circ}$ - $304^{\circ}$ & $229^{\circ}$ -
$328^{\circ}$  & $202^{\circ}$ - $349^{\circ}$ \\
\hline
\end{tabular}}
\caption{\label{tab1}The updated global fits of neutrino oscillation data has been presented  at 1$\sigma$, 2$\sigma$ and
3$\sigma$ CL. NO (IO) refers to normal (inverted)
neutrino mass ordering \cite{9}.}
\end{center}
\end{small}
\end{table}

\section{General Formalism}
The effective Majorana neutrino mass matrix $(M_{\nu})$ contains nine parameters comprising three neutrino masses ($m_{1}$, $m_{2}$, $m_{3}$), three mixing angles ($\theta_{12}$, $\theta_{23}$, $\theta_{13}$) and three CP violating phases ($\delta$, $\rho$, $\sigma$), can, in general, be expressed as

\begin{equation}\label{eq5}
M_{\nu}=UP\left(
\begin{array}{ccc}
    m_{1}& 0& 0 \\
  0 & m_{2} & 0\\
  0& 0& m_{3} \\
  \end{array}
  \right)P^{T}U^{T},
\end{equation}
 where $U$ denotes a $3\times3$ unitary matrix consisting of three flavor mixing angles ($\theta_{12}, \theta_{23}, \theta_{13}$), and one Dirac CP violating phase ($\delta$). $P=diag(e^{i\rho},e^{i\sigma},1)$, is a diagonal phase matrix containing two Majorana CP-violating phases ($\rho, \sigma$). The neutrino mass matrix $M_{\nu}$ can be  re-written as
\begin{equation}\label{eq6}
M_{\nu}=U\left(
\begin{array}{ccc}
    \lambda_{1}& 0& 0 \\
  0 & \lambda_{2} & 0\\
  0& 0& \lambda_{3} \\
\end{array}
\right) U^{T},
\end{equation}\\
where $\lambda_{1} = m_{1} e^{2i\rho},\lambda_{2} = m_{2} e^{2i\sigma} ,\lambda_{3} =
 m_{3}.$
For performing the present analysis, we consider the following parameterization of $U$  {used by Z. Xing} in \cite{23}:

\begin{small}
\begin{equation}\label{eq7}
U=\left(
\begin{array}{ccc}
 c_{12}c_{13}& s_{12}c_{13}& s_{13} \\
-c_{12}s_{23}s_{13}-s_{12}c_{23}e^{-i\delta} & -s_{12}s_{23}s_{13}+c_{12}c_{23}e^{-i\delta} & s_{23}c_{13}\\
 -c_{12}c_{23}s_{13}+s_{12}s_{23}e^{-i\delta}& -s_{12}c_{23}s_{13}-c_{12}s_{23}e^{-i\delta}& c_{23}c_{13} \\
\end{array}
 \right),
\end{equation}
\end{small}
where, $c_{ij} \equiv \cos \theta_{ij}$, $s_{ij}\equiv \sin \theta_{ij}$.\\
Using Eq. \ref{eq6}, any element $M_{ab}$ in the neutrino mass matrix can be expressed as
\begin{equation}\label{eq8}
M_{ab}=\sum_{i=1,2,3}U _{ai}U_{bi} \lambda_{i}.
\end{equation}

For texture two zero, we consider only two elements of $M_{\nu}$ to be zero, simultaneously. Hence we obtain two constraints from Eq. \ref{eq8}. Using them, we can derive the neutrino mass ratios ($\alpha, \beta$) in terms of mixing matrix elements \cite{24}. The three neutrino masses ($m_{1}, m_{2}, m_{3}$) can then be determined in terms of $\alpha, \beta$ as 
 \begin{equation}\label{eq9}
 m_{3}=\sqrt{\dfrac{\delta m^{2}}{\beta^{2}-\alpha^{2}}},  \qquad m_{2}=m_{3} \beta, \qquad m_{1}=m_{3} \alpha,
 \end{equation}
 where $\delta m^{2}$ denotes the solar neutrino mass squared difference.\\
 Using Eq. \ref{eq9}, we can express $\Sigma$ in terms of neutrino mass ratios as 
 \begin{equation}\label{eq10}
 \Sigma =\sqrt{\frac{\delta m^{2}}{\beta^{2}-\alpha^{2}}}(\alpha+\beta+1).
 \end{equation}
This relation can be used to deduce the $\Sigma$ as  a function of atmospheric mixing angle $\theta_{23}$. This is because  $\theta_{23}$ is most uncertain mixing angle in neutrino oscillation data.

 The  quasi degenerate masses of neutrinos implies either $m_{1} \lesssim m_{2}\backsim m_{3}$ for NO or $m_{3}\backsim m_{1} \lesssim m_{2}$ for IO. Following this assumption,  we can express  $|M_{ee}|$  explicitly in terms of $\Sigma$ as
 \begin{equation}\label{eq11}
 |M_{ee}|\simeq \dfrac{\Sigma}{3}|c_{12}^{2}c_{13}^{2}e^{2i\rho}+s_{12}^{2}c_{13}^{2}e^{2i\sigma}+s_{13}^{2}|.
 \end{equation}
 The general relations in Eqs. \ref{eq10} and \ref{eq11} will help to  understand the phenomenological results of the present analysis.
 
 \section{Numerical analysis}
 The experimental data regarding the  neutrino oscillation parameters at 1$\sigma$, 2$\sigma$ and 3$\sigma$ CL, respectively is given in Table \ref{tab1}].

For carrying out the analysis,  we scan the input neutrino oscillation parameters ($\theta_{12}, \theta_{23}$, $\theta_{13}$) and  the mass squared differences ($\delta m^{2}, \Delta m^{2}$) by allowing their random variation within their 3$\sigma$ CL ranges. The Dirac CP violating phase $\delta$ is allowed to vary from $0^{0}$ to $360^{0}$.  
Before discussing further, we,  again, emphasize here that all the seven texture zero cases are still allowed at 3$\sigma$ CL.

\begin{figure}[ht]
\begin{center}
\subfigure[]{\includegraphics[width=0.4\columnwidth]{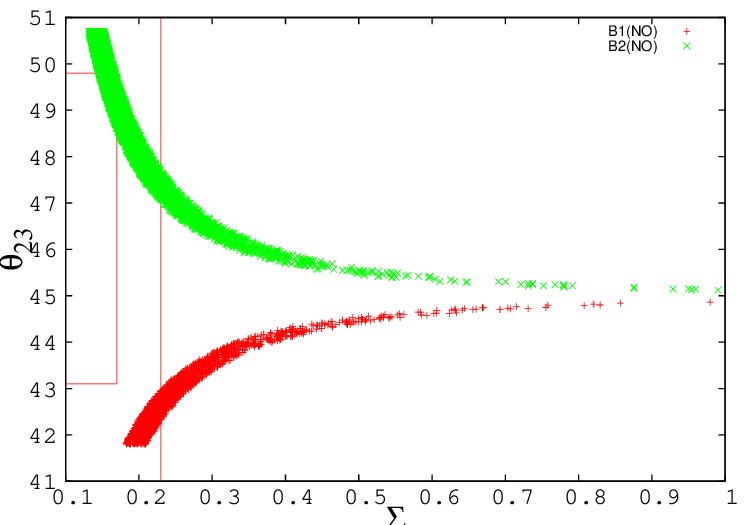}}
\subfigure[]{\includegraphics[width=0.4\columnwidth]{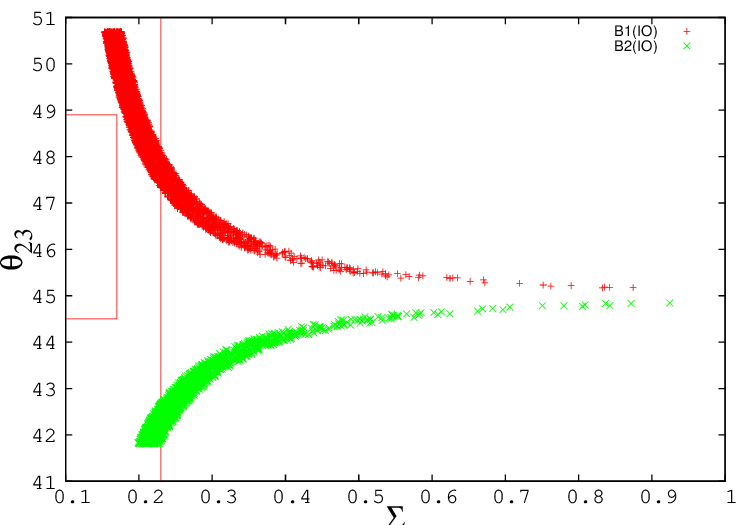}}\\

\caption{\label{fig1} Correlation plots between Sum of neutrino masses, $\Sigma$ (eV)  and  atmospheric mixing angle,  $\theta_{23}$ (degree), for  cases $B_{1}$ (red) and $B_{2}$ (green). The plots (a) and (b) correspond to NO and IO, respectively. The  box shows the   allowed parameter space at 2$\sigma$ CL, while solid line indicates the Planck's limit, $\Sigma=0.23$eV.}
\end{center}
\end{figure}

\begin{figure}[ht]
\begin{center}

\subfigure[]{\includegraphics[width=0.40\columnwidth]{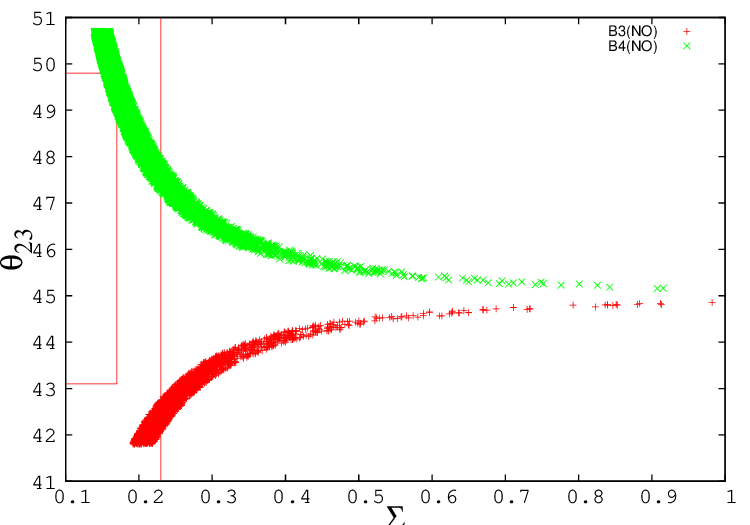}}
\subfigure[]{\includegraphics[width=0.40\columnwidth]{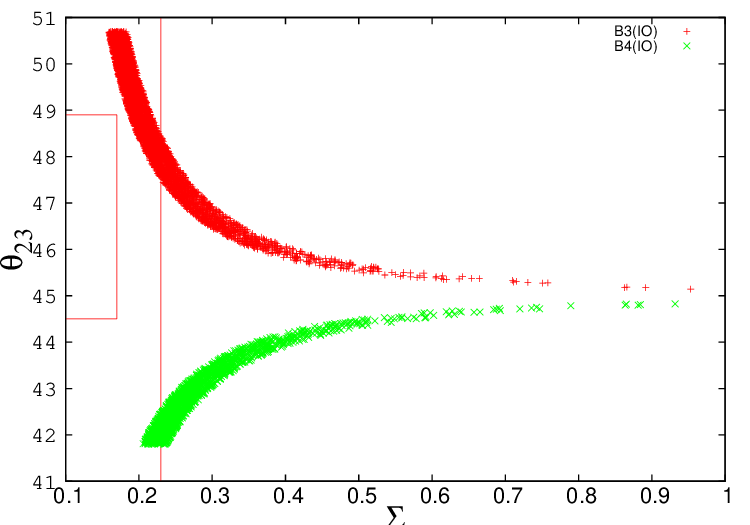}}\\
\caption{\label{fig2} Correlation plots between Sum of neutrino masses, $\Sigma$ (eV), and  atmospheric mixing angle,  $\theta_{23}$ (degree), for  cases $B_{3}$ (red) and $B_{4}$ (green). The plots (a) and (b) correspond to NO and IO, respectively. The  box shows the allowed parameter space at 2$\sigma$ CL, while solid line indicates the  Planck's limit, $\Sigma=0.23$eV.}
\end{center}
\end{figure}

\begin{figure}[ht]
\begin{center}
\subfigure[]{\includegraphics[width=0.40\columnwidth]{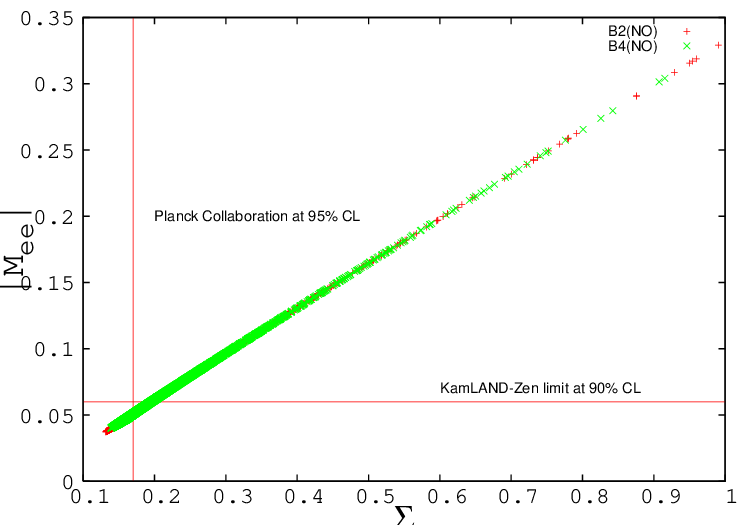}}
\subfigure[]{\includegraphics[width=0.40\columnwidth]{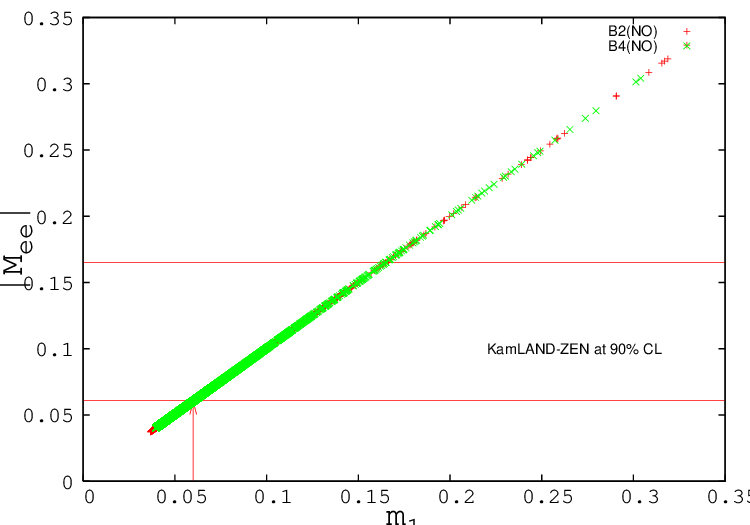}}

\caption{\label{fig3} (a) Correlation plots between Sum of neutrino masses, $\Sigma$ (eV),  and  effective mass term,  $|M_{ee}|$ (eV), for  cases $B_{2}$ (red) and $B_{4}$ (green) for NO.  The vertical red line indicates the recent Planck's collaboration limit, while horizontal line indicates the recent KamLAND-ZEN limit at 90$\%$  CL.  (b) Correlation plots between lightest neutrino mass, $m_{1}$ (eV),  and  effective mass term,  $|M_{ee}|$ (eV), for  cases $B_{2}$ (red) and $B_{4}$ (green) for NO. The solid bold line indicates the region for recent KamLAND-ZEN limit at 90 $\%$ CL, and vertical arrow points towards the upper bound on $m_{1}$.  }
\end{center}
\end{figure}
\begin{figure}[ht]
\begin{center}

\subfigure[]{\includegraphics[width=0.40\columnwidth]{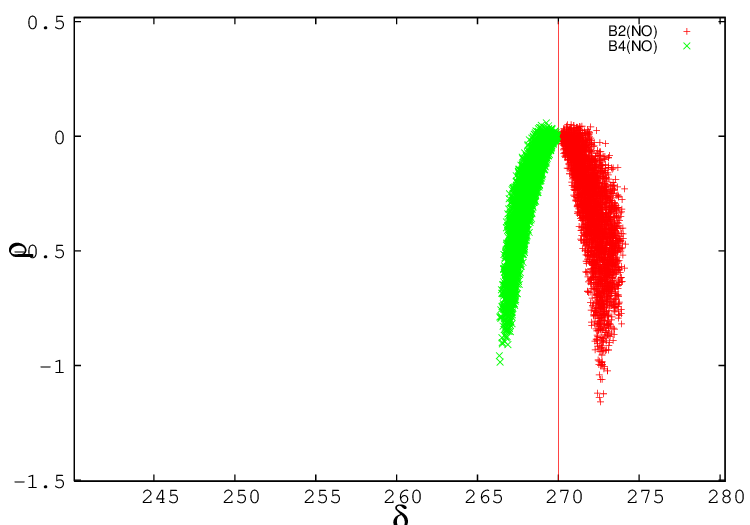}}\\
\subfigure[]{\includegraphics[width=0.40\columnwidth]{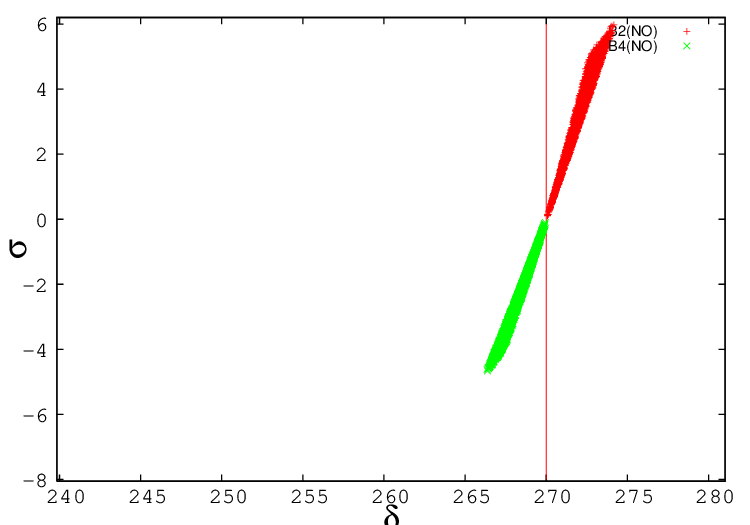}}\\

\caption{\label{fig4}  Correlation plots between Majorana phases, $\rho$, $\sigma$ (degree) and Dirac CP-violating phase, $\delta$ (degree), for  cases $B_{2}$ (red) and $B_{4}$ (green) for NO, respectively.    }
\end{center}
\end{figure}

\begin{figure}[ht]
\begin{center}
\subfigure[]{\includegraphics[width=0.40\columnwidth]{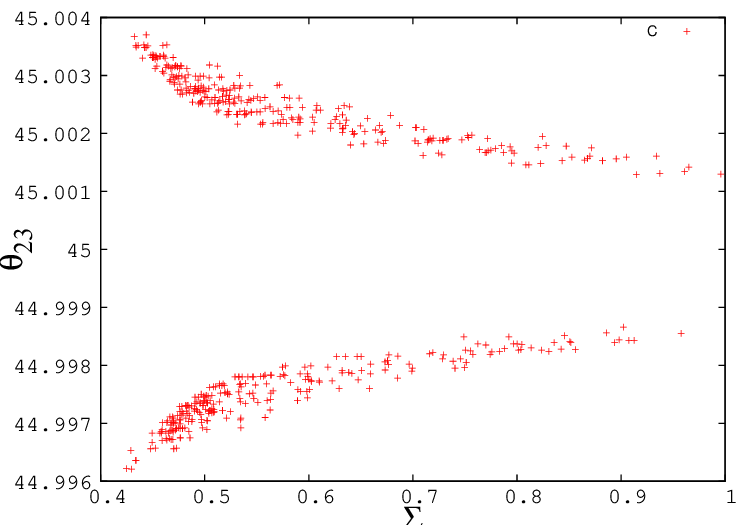}}
\subfigure[]{\includegraphics[width=0.40\columnwidth]{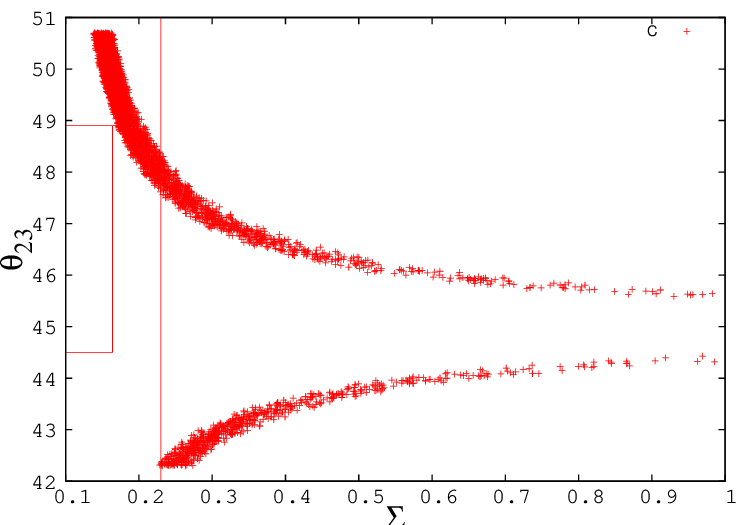}}\\

\caption{\label{fig5} Correlation plots between Sum of neutrino masses, $\Sigma$ (eV)  and  atmospheric mixing angle,  $\theta_{23}$ (degree), for  case C for (a) NO (b)IO. The  box indicates the allowed parameter space at 2$\sigma$ CL. The vertical line indicates the Planck's limit, $\Sigma=0.23$eV.  }
\end{center}
\end{figure}

 In the following discusion, we explicate the phenomenological results through the correlation  of $\Sigma$ and $\theta_{23}$ for  the  texture two zero cases. For the sake of comparison, we also provide the result at relatively higher limit, $\Sigma<0.23$ eV \cite{32}, as used by D. Meloni in \cite{25}.  The correlation plots of $\Sigma$ and $\theta_{23}$ for all the five viable cases, have been complied in Figs. \ref{fig1}, \ref{fig2}, \ref{fig5}.
 
 Further, we provide the approximate analytical relations of $\Sigma$, as a function of $\theta_{23}$ using Eq.\ref{eq10}.  With the help of these expressions, we attempt to show analytically why a particular case is ruled out.
   To understand the implication of Planck's limit on 0$\nu\beta\beta$ decay, we   relate the $|M_{ee}|$ with $\Sigma$ analytically for the viable cases.  The details of the analysis are presented as follows:
  
  For cases $B_{1}$ and $B_{3}$, the approximate neutrino masses ($m_{1}, m_{2}, m_{3}$) can be expressed  in the leading order of $s_{13}$ as \cite{23}  
 \begin{equation}\label{eq12}
m_{1}\simeq m_{2} \simeq m_{3}tan^{2}\theta_{23},
 \end{equation}
 \begin{equation}\label{eq13}
 m_{3}\simeq \sqrt{\frac{\delta m^{2}}{1-tan^{4} \theta_{23} }}.
  \end{equation}
 { From Eq.(\ref{eq12}), one can conclude that for  $\theta_{23}> 45^{0}$ ($\theta_{23}< 45^{0}$), IO(NO) is allowed.  Using Eq.(\ref{eq13}), as $\theta_{23}$ approaches to $45^{0}$, the three neutrino masses, $m_{1}, m_{2}$ and $m_{3}$ tend  to infinite value, respectively. Hence, the recent cosmological scale  rules out $\theta_{23}=45^{0}$ as also evident in Fig.[\ref{fig1}, \ref{fig2}.  }
  
   Using Eqs. \ref{eq12} and \ref{eq13},  we can find $\Sigma$ in terms of $\theta_{23}$ in the leading order term 
    
  \begin{equation}\label{eq14}
  \Sigma \simeq \sqrt{\frac{\delta m^{2}}{1-tan^{4} \theta_{23} }} (2tan^{2} \theta_{23}+1).
  \end{equation}

 Using the 2$\sigma$ range of neutrino oscillation data [Table \ref{tab1}], we find, $\Sigma_{min}\simeq 0.24$eV,  for NO. Hence, cases $B_{1}$ as well as $B_{3}$ are, now, ruled out for NO. The observation can be validated from Figs. \ref{fig1}(a) and \ref{fig2}(a), showing the correlation  between $\Sigma$ and $\theta_{23}$ for cases $B_{1}$ and $B_{3}$, respectively. On the other hand, for IO,  $\Sigma_{min} \simeq 0.22$eV, which remain excluded in the scenario of present limit on $\Sigma$.   In Figs. \ref{fig1}(b) and \ref{fig2}(b), we have shown the correlation between $\Sigma$ and $\theta_{23}$ for cases $B_{1}$ and $B_{3}$,  respectively, for IO. It is explicitly shown that the allowed parameter space is excluded by the present limit on $\Sigma$.  On the contrary,  the  parameter space for cases $B_{1}$ and $B_{3}$, respectively, lies within the region of experimental bound,  $\Sigma<0.23$eV. Hence  cases $B_{1}$ and $B_{3}$ are now ruled out for both NO and IO at 2$\sigma$ level.
 
 Similarly,  for cases $B_{2}$ and $B_{4}$, we have presented the expressions for $m_{1}, m_{2}$ and $m_{3}$ in the leading order of $s_{13}$ as \cite{23}
 \begin{equation}\label{eq15}
m_{1}\simeq m_{2} \simeq m_{3}cot^{2}\theta_{23},
 \end{equation}
 \begin{equation}\label{eq16}
 m_{3}\simeq \sqrt{\frac{\delta m^{2}}{1-cot^{4} \theta_{23} }}.
  \end{equation}
  Similar to cases $B_{1}$ and $B_{3}$,  $\theta_{23}$ = $45^{0}$ is ruled out for cases $B_{2}$ and $B_{4}$ as well, while  $\theta_{23}> 45^{0}$ ($\theta_{23}< 45^{0}$) leads to NO (IO). 
 Using Eqs. \ref{eq15} and \ref{eq16}, we can write 
\begin{equation}\label{eq17}
  \Sigma \simeq \sqrt{\frac{\delta m^{2}}{tan^{4} \theta_{23}-1 }} (2+tan^{2} \theta_{23}).
  \end{equation}

Taking into account the 2$\sigma$ range of neutrino oscillation parameter [Table \ref{tab1}], we find, $\Sigma_{min} \simeq 0.154$eV for NO, which lies in the range of Planck's limit. In Fig. \ref{fig1}(a), we  present the correlation between  $\Sigma$ and $\theta_{23}$ for case $B_{2}$ for NO.   On comparing,  we find that although both the cosmological limits allow the parameter space of $\theta_{23}$, however for $\Sigma<0.17$eV, the parameter space for the same is now reduced to an appreciable extent. More explicitly, $\theta_{23}<45^{0}$ is disallowed for NO. 

Using 2$\sigma$ CL of  neutrino oscillation data, we get,  $\Sigma_{min} \simeq 0.24$eV from Eq.\ref{eq17}, for IO, which clearly lies outside the present Planck's limit. Similar phenomenological results can be obtained for case $B_{4}$ for both NO and IO. The correlation plots have been complied in Figs. \ref{fig1}, \ref{fig2} for the sake of completion.
Earlier, D. Meloni et al. \cite{25} have also pointed out the similar predictions for cases $B_{2}$ and $B_{4}$,  respectively, at 1$\sigma$ CL. In comparison with their results, our numerical result is compatible at relatively higher CL, and also the available parameter space of $\theta_{23}$ for NO, is more constricted in our analysis.
  
Mathematically, one can relate the effective mass $|M_{ee}|$ in terms of $\Sigma$ and $\theta_{23}$ using Eq.\ref{eq11} 
\begin{equation}\label{eq18}
|M_{ee}|\simeq \dfrac{\Sigma}{3} tan^{2} \theta_{23}.
\end{equation}

The above relation has been obtained in the leading order term of $\theta_{13}$. Using the best fit of $\theta_{23}$, $\theta_{23} =47.7^{0}$   we find, $|M_{ee}|<0.0684$eV. This  bound for $|M_{ee}|$ is  found to be very close to the senstivity limit obtained from KamLAND-Zen experiment, (i.e. $|M_{ee}| \leq 0.06$eV  \cite{16}).   The strong correlation between $\Sigma$ and $ |M_{ee}|$ also verifies this prediction on $|M_{ee}|$ [Fig. \ref{fig3}(a)].  Hence, our result  for case $B_{2}$ resonates with the  experimental bound of KamLAND-Zen experiment.  In Fig.\ref{fig3} (b), a strong linear correlation between effective mass $|M_{ee}|$ and lightest neutrino mass, $m_{1}$, have been shown indicating the quasi-degenerate spectrum for cases $B_{2}$ and $B_{4}$, respectively . It is clear that significantly larger part of the parameter space above the sensitivity limit of KamLAND-Zen experiment is, now, ruled out. Fig.\ref{fig3} (b) also put the bound on lightest neutrino mass,  $m_{1}\lesssim 0.06$eV for cases $B_{2}$ and $B_{4}$, respectively.

 In Table \ref{tab1},  $\delta$ ranges from $157^{0}$ to $349^{0}$ for NO, and  $202^{0}$ to $349^{0}$ for IO  at 3$\sigma$ CL. This new result automatically excludes the parameter space around $\delta \simeq 90^{0}$,  and allows $\delta\simeq 270^{0}$ for cases $B_{2}$ and $B_{4}$. In Figs. \ref{fig4}(a),(b), as $\delta$ is approaching to $270^{0}$, Majorana phases ($\rho, \sigma$) also approach to vanishing value, otherwise $\rho \simeq 0^{0}, \sigma \simeq 0^{0}$ even at 3$\sigma$ CL.
  In comparison with the earlier analyses \cite{23,24,25,26,27} , our results for $\delta$  overlaps with the recent global fits on $\delta$ \cite{14}, and also holds true at significant CL.  

In the present scenario of analysis,  the viable cases  $B_{2}$ and $B_{4}$  seem to be very interesting from the experimental point of view since they  simultaneously incorporate the updated T2K, Super-Kamiokande and NO$\nu$A results. The combined results favor the NO ( $m_{1} < m_{2}<m_{3}$),  $\delta \simeq 270^{0}$ and $\theta_{23} > 45^{0}$  at higher CL.  Earlier in Ref. \cite{26}, S. Zhou  also discusses the similar predictions for these cases along with the cases $A_{1}$ and $A_{2}$, respectively at 1$\sigma$ CL.

 The phenomenological implications of $B_{2}$ and $B_{4}$ are almost similar  regarding  the neutrino mass ordering, octant of $\theta_{23}$  and Dirac CP-violating phase ($\delta$), therefore it seems to be difficult to differentiate them  at low energies. 
To this end,  we have to measure the CP-violating
phase as precisely as possible, e.g., an accuracy less than $1^{0}$
is required if $\delta \simeq 270^{0}$ is confirmed. Further, it may be possible that the phenomenological predictions for these cases come out to be different for the flavor models at high energy scale.
 
For case C, we can rewrite the expressions for $\Sigma$, as presented in \cite{25} 
 \begin{equation}\label{eq19}
 \Sigma = \sqrt{\Delta m^{2}} \frac{[3-s_{13}^{2}(2-cos2\delta)]}{2s_{13}|cos\delta|},
 \end{equation}
  for NO if maximal $\theta_{23}$ is considered, while for IO,
  \begin{small}
   \begin{eqnarray}\label{eq20}
 &&\Sigma \simeq \frac{\sqrt{\delta m^{2}}}{2} \bigg [\frac{2}{s^{2}_{12}}-\frac{2s_{13}c_{12}t_{2(23)}}{s_{12}}\bigg(-1+\frac{c_{\delta}}
{s^{2}_{12}}\bigg) 
 +s^{2}_{13}\bigg( \frac{(-2c_{\delta}c^{2}_{12}+1)}{s^{2}_{12}}+\frac{c^{2}_{\delta}(3c^{2}_{12}-s^{2}_{12})t^{2}_{2(23)}}{s^{4}_{12}}\bigg)\bigg],
 \end{eqnarray}
 \end{small}
 where $t_{2(23)}\equiv tan2\theta_{23}$, $c_{\delta}\equiv cos \delta$.
The analysis done by D. Meloni et al. \cite{25} excludes the NO for case C at 1$\sigma$ CL, which remains consistent with our result as well [Fig. \ref{fig5} (a)]. Using the 2$\sigma$ range of  neutrino oscillation parameters, from Eq.(\ref{eq20}), we find, $\Sigma_{min} \simeq 0.174 $eV, which marginally lies outside the current Planck's limit, $\Sigma<0.17$eV. This indicates that case C is now  ruled out for IO as well. This result is different from ref \cite{25}, which allows IO  for case C.    To comprehend this  result, we have presented the comparison, considering the  Planck's limit, $\Sigma<0.17$eV and $\Sigma<0.23$eV  respectively [Figs. \ref{fig5} (b)].   The correlation between $\Sigma$ and $\theta_{23}$  clearly excludes the allowed parameter space for  $\Sigma<0.17$eV , owing to the present refinement in $\theta_{23}$ and $\Sigma$ at 2$\sigma$ CL.
  
It must be noted that, even at $3\sigma$ CL, the parameter space of all the five cases are found to be tightly constrained under the current experimental scenario as evident from Figs.\ref{fig1},\ref{fig2},\ref{fig3},\ref{fig4},\ref{fig4},\ref{fig5}, however no specific conclusion can be made regarding the physical parameters. In Table \ref{tab2}, we have encapsulated the current status of texture two zero cases as well as the predictions on $\Sigma$, $|M_{ee}|$, $\delta$ in light of latest experimental data at 2$\sigma$ CL. 
 
\begin{table}[ht]
\begin{small}
\begin{center}

\begin{footnotesize}
\resizebox{14cm}{!}{
\begin{tabular}{|c|c|c|c|c|c|c|c|c|}
  \hline
Cases&\multicolumn{4}{c|}{
NO}&\multicolumn{4}{c|}{
IO} \\ \hline 
 &  Octant & $\delta$ &$|M_{ee}|$ &$\Sigma$& Octant & $\delta$ &$|M_{ee}|$ &$\Sigma$
\\ 
\hline 

$B_{1}$  &
$\times$& $\times$& $\times$&  $\times$&
$\times$& $\times$& $\times$ & $\times$ \\ 
 \hline
 $B_{2}$  &
$\theta_{23} >45^{0}$& $270.05^{0}-273.2^{0}$& $|M_{ee}|>0.0389eV$ &  $\Sigma >0.141eV $&
$\times$& $\times$& $\times$ & $\times$ \\ 
 \hline
 $B_{3}$  &
$\times$& $\times$& $\times$&  $\times$&
$\times$& $\times$& $\times$ & $\times$ \\ 
 \hline
 $B_{4}$  &
$\theta_{23} >45^{0}$& $267.25^{0}-270^{0}$& $|M_{ee}|>0.0422eV$ &  $\Sigma >0.151eV $&
$\times$& $\times$& $\times$ & $\times$ \\ 
 \hline
 $C$  &
$\times$& $\times$& $\times$&  $\times$&
$\times$& $\times$& $\times$ & $\times$ \\ 
 \hline
 
\end{tabular}}
\caption{\label{tab2} Predictions regarding the current status of five viable cases alongwith  the allowed parameter space of  octant of $\theta_{23}$, Dirac CP-violating phase ($\delta$), effective neutrino mass ($|M_{ee}|$), and sum of neutrino masses ($\Sigma$) at 2$\sigma$ CL.}
\end{footnotesize}
\end{center}
\end{small}
\end{table}

\section{Summary and Conclusions}
To summarize our discussion, we have presented the complete analysis of texture two zero Majorana mass matrices in the light of the combined results of latest neutrino oscillation experiments, Planck's collaboration  as well as KamLand-Zen experiments. For the analysis, we have considered only five out of seven viable cases, among them, only two cases ($B_ {2} $, and $B_ {4} $) with NO, are found to be consistent with the combined data, while remaining three cases ($B_ {1}, B_ {3} $ and C) are now ruled out for both NO and IO at 2$\sigma$ CL.  
 More importantly, the phenomenological results of the viable cases remain in tune with current observations of T2K, Super-Kamiokande and NO$\nu$A experiments. In addition, for viable cases, the predictions for effective mass term $|M_{ee}| $ is also found to be consistent with the latest KamLAND Zen limit on the neutrino less double beta decay.
 In future, the precise measurement of $\delta$  serves as an important discriminator for the  viable  two-zero textures.

To conclude our discussion, we would like to say that these predictions are exciting as far as the present experimental scenario is concerned. The further progress and precision in the statistical data from long baseline experiments, cosmological as well as KamLand-Zen experiments could help us to validate these predictions regarding the texture two zero at a significantly higher confidence level. This, in consequence, can  give us a new insight into the structure of lepton mass matrices,  and  possibly helps us to throw some light on the origin of neutrino masses and dynamics of flavor mixing and leptonic CP violation.

\section*{Data availability}

The data used to support the findings of this study is included within this article .

\section*{Conflicts of Interest}
The author declares that there are no conflicts of interest regarding the publication of this paper.

\section*{Acknowledgment}
The author would like to thank the Principal of M. N. S. Government College,  Bhiwani, Haryana, India, for providing the necessary facilities to work. \\


\begin{thebibliography}{43}

\bibitem{1}  K. Abe et al. [T2K collaboration], Phys. Rev. Lett. 107, 041801 (2011),
     arXiv: 1106.2822 [hep-ex].

\bibitem{2}  P. Adamson et al. [MINOS collaboration], Phys. Rev. Lett. 107, 181802 (2011),
     arXiv: 1108.0015 [hep-ex].

\bibitem{3}  Y. Abe et al., [Double Chooz collaboration], Phys. Rev. Lett. 108, 131801 (2012),
     arXiv: 1112.6353 [hep-ex].

\bibitem{4}  F. P. An et al., [Daya Bay collaboration], Phys.  Rev. Lett. 108, 171803 (2012),
     arXiv: 1203.1669 [hep-ex].

\bibitem{5}  Soo-Bong Kim, for RENO collaboration, Phys. Rev. Lett. 108, 191802 (2012),
     arXiv: 1204.0626[hep-ex].
 \bibitem{6}   G. Bellini et al., Phys. Rev. Lett. 107 (2011) 141302, arXiv:1104.1816 [hep-ex].
 \bibitem{7} P. Adamson et al. (NOvA), Phys. Rev. Lett. 118, 231801 (2017), arXiv:1703.03328 [hep-ex].
 
 \bibitem{8} P. F. de Salas, D. V. Forero, C. A. Ternes, M. Tortola, J. W. F. Valle, Phys. Lett. B, 782, 633-640 (2018), arXiv:1708.01186 [hep-ph].

 \bibitem{9} F. Capozzi, E. Lisi, A. Marrone, A. Palazzo,  arXiv: 1804.09678 [hep-ph].
 \bibitem{10}  I. Esteban,  M. C. Gonzalez-Garcia, A. Hernandez-Cabezudo, et al., JHEP 01, 106 (2019),  arXiv: 1811.05487 [hep-ph].
 
 \bibitem{11} A. Radovic, "Latest Neutrino Oscillation Results from NOvA," seminar at Fermilab (12
January 2018), available at https://www-nova.fnal.gov. See also the subsequent NOvA talks, up to March
2018.
 \bibitem{12} "T2K and NOvA collaborations to produce joint neutrino oscillation analysis by 2021."
Full announcement (30 January 2018) available at http://t2k-experiment.org/2018/01/
t2k-nova-announce/
 \bibitem{13} K. Abe et al. [T2K Collaboration],  Phys. Rev. D 96 (2017) no.9, 092006, arXiv: 1707.01048 [hep-ex].
 \bibitem{14} Super-Kamiokande Collaboration: $\chi^{2}$ maps derived from full 3$\sigma$ oscillations analysis,  http://www-sk.icrr.u-tokyo.ac.jp/sk/publications/result-e.html.

\bibitem{15} P. A. R. Ade et al., [Planck Collaboration], Astronomy and Astrophysics 594, A13 (2016),  arXiv: 1502.01589 [astro-ph.CO].
\bibitem{16} KamLAND-Zen Collaboration, Phys. Rev. Lett., 117, 082503 (2016), arXiv: 1605.02889.
\bibitem{17}  J. B. Albert et. al [Exo-200 Collaboration], Nature 510, 229-234 (2014), arXiv: 1402.6956 [nucl-ex].
\bibitem{18} S. Dell'oro, S. Marcocci, M. Viel and F. Vissani, Adv. High Energy Phys., 2162659 (2016), arXiv: 1601.07512.
\bibitem{19} W. Rodejohann, Int. J. Mod. Phys. E, 20, 1833 (2011), arXiv: 1106.1334 [hep-ph].
\bibitem{20} F. T. Avignone III, S. R. Elliott, J. Engel, Rev. Mod. Phys. 80, 481 (2008), arXiv: 0708.1033 [nucl-ex]; J. J. Gomez-Cadenas, J. Martin-Albo, M. Mezzetto, F. Monrabal, M. Sorel,  Riv. Nuovo  Cim. 35 (2012) 29-98, arXiv: 1109.5515 [hep-ex].
\bibitem{21}  M. Gupta, G. Ahuja, Int. J. Mod. Phys. A 26, 2973 (2011) and references
therein.
 
\bibitem{22} Rohit Verma, Advances in High Energy Physics, Volume 2016, Article ID 2094323, arXiv: 1607.00958 [hep-ph].
  
  
\bibitem{23} Paul H. Frampton, Sheldon L. Glashow and Danny Marfatia, Phys. Lett. B 536, 79 (2002), hep-ph/0201008; Zhi-zhong Xing, Phys. Lett. B 530, 159 (2002), hep-ph/0201151.
\bibitem{24}  H. Fritzsch, Z. Z. Xing, S. Zhou, JHEP 1109, 083 (2011), arXiv: 1108.4534v2 [hep-ph].
 \bibitem{25} D. Meloni, A. Meroni, E. Peinado, Phys. Rev. D 89 (2014) 053009, arXiv: 1401.3207 [hep-ph].
 \bibitem{26} Shun Zhou, Chin. Phys. C 40 (2016) no.3, 033102, arXiv: 1509.05300 [hep-ph].
\bibitem{27} Madan Singh, Gulsheen Ahuja, Manmohan Gupta,  Prog. Theor. Exp. Phys. (PTEP) 
         2016 (12): 123B 08, arXiv: 1603.08083 [hep-ph].
\bibitem{28} P. H. Frampton, M. C. Oh and T. Yoshikawa, Phys. Rev. D 66, 033007 (2002).
\bibitem{29} Walter Grimus and Luis Lavoura, J. Phys. G 31, 693-702 (2005).
\bibitem{30} W. Grimus, A.S. Joshipura, L. Lavoura, and M. Tanimoto, Eur. Phys. J. C 36, 227 (2004).
\bibitem{31} H. C. Goh, R. N. Mohapatra and Siew-Phang Ng, Phys. Rev. D 68, 115008 (2003). 
\bibitem{32} P. A. R. Ade et al. [Planck Collaboration], arXiv: 1303.5076 [astro-ph.CO].


\end{thebibliography}
\end{document}